\documentclass[
    ,final            
  ]
  {aipproc}

\layoutstyle{6x9}
\pdfoutput=1

\begin{document}

\title{Reflective Scattering and Unitarity}


\author{S.M. Troshin}{}

\author{N.E. Tyurin}{
  address={Institute for High Energy Physics, Protvino, Moscow Region, 142281, Russia}
}

\begin{abstract}
Interpretation of unitarity saturation as reflective
scattering is discussed.  Analogies  with optics and Berry phase
 alongside with
the experimental consequences of the proposed interpretation  at the
LHC energies are considered.
\end{abstract}

\maketitle


\section{Unitarity: absorptive vs reflective scattering}
 The essential and persisting
problems of  QCD  are related to confinement and spontaneous
chiral symmetry breaking phenomena. It is the field of
collective, coherent interactions of quarks and gluons resulting
in formation of the asymptotic states --- the colorless,
experimentally observable particles.
Closely related --- the total
cross--section  growth with energy --- constitutes one of the
important unanswered question in the theory.
 Elastic scattering gives significant contribution
to the total cross-section, which  by means of optical theorem,
 is related to the forward elastic scattering amplitude. Thus, the process of
 elastic scattering, where all hadron constituents interact
coherently,  can  serve as a tool in the confinement studies.
Relative strength of elastic and inelastic processes is regulated
by unitarity.  It is important to note here that unitarity is
formulated for the asymptotic colorless hadron on-mass shell
states and is not directly connected to the fundamental fields of
QCD --- quarks and gluons. The same is valid for the analyticity
which is relevant for the  scattering amplitudes of the observable
particles only. It is not clear
what these fundamental principles
 imply for the confined objects.

Unitarity or conservation of probability, which can be   written in terms
of the scattering matrix as following
\begin{equation}\label{ss}
SS^+=1,
\end{equation} implies an
existence at high energies of the two scattering modes - shadow one (absorptive scattering)
 and antishadow (reflective scattering).
  Indeed, writing unitarity relation for the partial wave amplitudes $f_l(s)$ :
\begin{equation}\label{ul}
\mbox{Im} f_l(s)=|f_l(s)|^2+\eta_l(s),
\end{equation}
where elastic $S$-matrix is related to the amplitude as
\begin{equation}\label{sl}
S_l(s)=1+2if_l(s)
\end{equation}
 and $\eta_l(s)$ stands for the contribution of the intermediate inelastic channels to the
elastic scattering with the orbital angular momentum $l$,
we can easily observe that the relation (\ref{ul}) turns out to be a
 quadratic equation in the case of the pure imaginary
scattering amplitude and the elastic amplitude  appears  to be  not a
singe-valued function of $\eta_l$. But, only one of the two solutions of the equation corresponding
to the relation (\ref{ul})
 is considered almost everywhere:
\begin{equation}\label{us}
f_l(s)=\frac{i}{2}(1-\sqrt{1-4\eta_l(s)}),\quad \mbox{i.e.} \quad |f_l|\leq 1/2,
\end{equation}
while another one:
\begin{equation}\label{uas}
f_l(s)=\frac{i}{2}(1+\sqrt{1-4\eta_l(s)}),\quad \mbox{i.e.} \quad 1/2 \leq |f_l|\leq 1
\end{equation}
is neglected.
However, there are no reasons for its neglecting at small and moderate values of orbital
angular momentum $l$ \cite{intja}.
The two above mentioned solutions can be easily reconciled in the uniform way in the
$U$-matrix unitarization approach \cite{umat}, which represents elastic $2\to 2$ scattering
matrix $S(s,b)$ in the impact parameter picture in the following form
\begin{equation}
S(s,b)=\frac{1+iU(s,b)}{1-iU(s,b)}. \label{um}
\end{equation}
 $U(s,b)$ is the generalized reaction matrix, which is considered to be an
input dynamical quantity. The transform (\ref{um}) is one-to-one and easily
invertible. Inelastic overlap function $\eta(s,b)$
can also be expressed through the function $U(s,b)$ by the relation
\begin{equation}
\eta(s,b)=\frac{\mbox{Im} U(s,b)}{|1-iU(s,b)|^{2}}\label{uf},
\end{equation}
and the only condition to obey unitarity
 is $\mbox{Im} U(s,b)\geq 0$.

 In what follows we consider for simplicity
 the case of pure imaginary $U$-matrix and make the replacement $U\to iU$.
 The value of energy corresponding to the full absorption
at central collisions  $S(s,b)|_{b=0}=0$
will be denoted as $s_0$ and it is determined by the  equation
$U(s,b)|_{b=0}=1$.
In the energy region $s\leq s_0$ the scattering in the whole range of impact parameter variation
has a shadow nature and correspond to solution (\ref{us}), the $S$ matrix varies in the range
$0\leq S(s,b)<1$. But when the energy is higher than the threshold
value $s_0$, the scattering picture at small values of impact parameter $b\leq R(s)$
corresponds to the solution Eq. (\ref{uas}), where $R(s)$
is determined by solution of equation $S(s,b=R)=0$.
The $S$-matrix variation region is then $-1<S(s,b)\leq 0$ at $s\geq s_0$
and $b\leq R(s)$.
\section{Analogies with optics and Berry phase}
There is a close analogy
here with the light reflection off a dense medium, when the phase of the reflected light is changed by
$180^0$. Therefore, using the optical concepts \cite{gottfr}, the above behavior
 of $S(s,b)$ should be interpreted
as an appearance of a reflecting ability of scatterer due to increase of
 its density beyond some critical value, corresponding to refraction index  noticeably
 greater than unity. In another words, the scatterer has now not only
 absorption ability (due to  presence of inelastic channels), but it starts to be reflective at very
 high energies and its central part ($b=0$) approaches to the completely
 reflecting limit ($S=-1$) at $s\to\infty$.
In order to combine shadowing at large values of $b$ with antishadowing
  in central collisions the real part of the phase shift should have the  dependence
\[
\delta_R (s,b)=\frac{\pi}{2}\theta(R(s)-b).
\]

Such a behavior of $ \delta_R (s,b)$ just takes place in the $U$-matrix form of unitarization.
Indeed, the phase shift $\delta (s,b)$
 can be expressed in terms of the function $U(s,b)$ as following
 \begin{equation}\label{del}
 \delta (s,b)=\frac{1}{2i}\ln\frac{1-U(s,b)}{1+U(s,b)}.
\end{equation}
It is clear that in the region $s>s_0$ the function
 $\delta (s,b)$ has a real part $\pi/2$ in the region $0<b\leq R(s)$, while $\delta_I (s,b)$ goes to
  infinity at $b=R(s)$.

It also leads to another interesting similarity, namely it  allows one to consider $ \delta_R $
 as an analog of the geometric Berry phase in quantum mechanics which appears
 as a result of a cyclic time evolution of the Hamiltonian parameters \cite{berry}.
This interesting phenomenon  can be observed in many physical systems
and is, in fact, a feature  of a system that depends only on the path it evolves along of.
 In the case of pure imaginary elastic scattering amplitude the contribution
of the inelastic channels $\eta$ can be considered as a
 parameter which determines  due to unitarity (but not in a unique way)  the elastic $S$-matrix.
 We can  vary variable $s$ (and/or $b$) in a way
 that the parameter $\eta$ (which has a peripheral $b$-dependence)
  evolves  cyclically from $\eta_i<1/4$ to $\eta_{max}=1/4$ and
  again to the value $\eta_f$, where $\eta_f=\eta_i$ (loop variation).
  As a result
 the  non-zero phase appears ($\delta_R=\pi/2$ at $b\leq R(s)$) and this phase
  is independent of the details
   of the energy evolution (Fig. 1).
\begin{figure}[hbt]
\resizebox{12cm}{!}{\includegraphics*{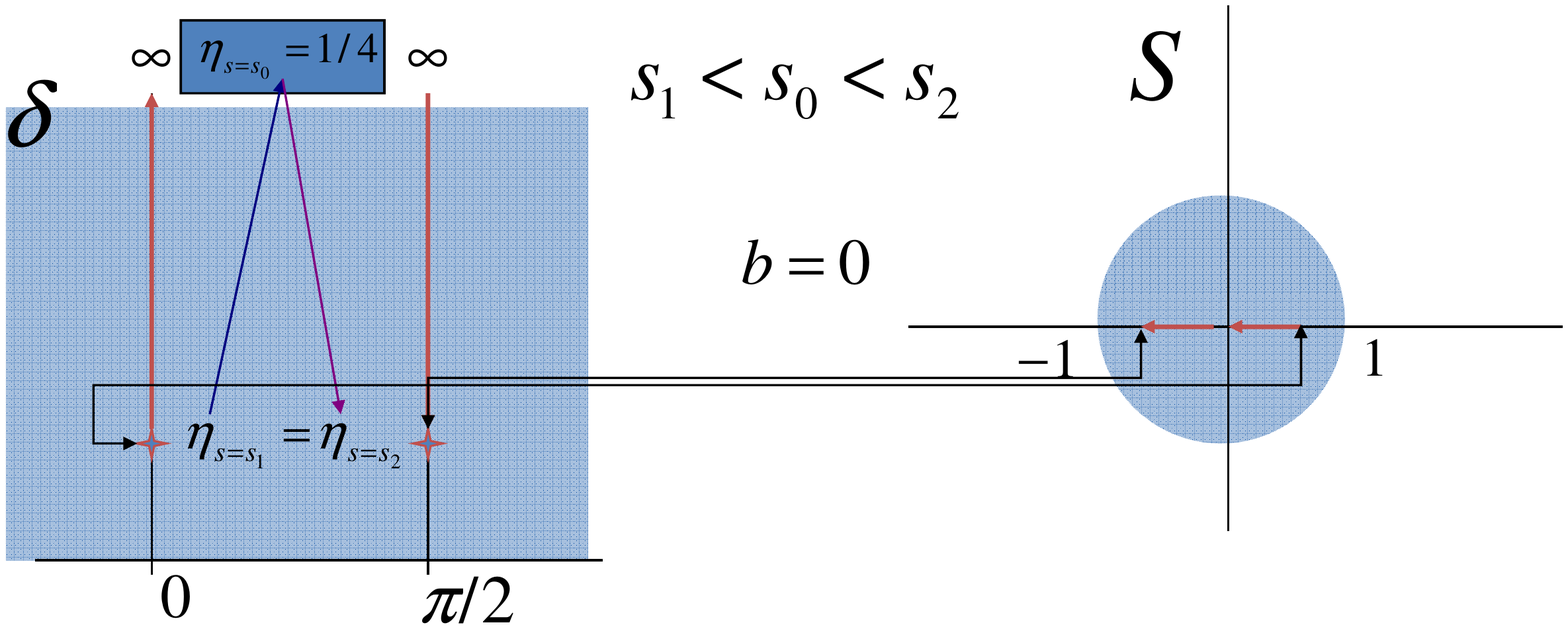}}
\caption{Variation of the scattering phase $\delta$ with energy}
\end{figure}
Since the Berry phase has a geometrical  origin  it is tempting to relate the
step energy dependence of $\delta_R(s)$ with existence of an extra dimension at TeV
energy scale.

Thus,  we can summarize  that
the physical scattering picture beyond the black disk limit  evolves with energy
by simultaneous  increase of the reflective ability (i.e. $|S(s,b)|$ increases
with energy and  $\delta_R=\pi/2$),
  and decrease of the absorptive ability $1-|S(s,b)|^2$ at the impact parameters $b<R(s)$.

Having in mind a quark-gluon  structure of hadrons it would be interesting to find a particular
 microscopic  mechanisms  related
 to  the collective quark-gluon dynamics in head-on collisions which can be envisaged as an
 origin of the reflection phenomenon.
 One can  speculate at this point and  relate the appearance of the reflective scattering
  to the Color-Glass Condensate in QCD (cf. e.g. \cite{mcler} and references therein) merely ascribing
  the reflective ability to Glazma.

The particular model for $U$-matrix \cite{prev} allows to give a rather good
description of the observables in elastic hadron scattering and provide relevant predictions
for them at the LHC energies.
It is interesting that power-like dependence of the differential cross-sections at large angles
appeared to be related to the total cross-section growth. Transition to the reflective scattering
 regime is also responsible for the existence of the knee in the energy spectrum of cosmic rays.
There is an interesting possibility that the reflective scattering mode could be discovered
at the LHC by measuring $\sigma_{el}/\sigma_{tot}$ ratio which would be
greater than the black disk  value $1/2$ \cite{bbl}.
However,
the asymptotical regime
is expected in the model at $\sqrt{s}> 100$ $TeV$ only.
Increasing weight of the  reflective scattering at the LHC energies
would lead to  the less prominent dip-bump structure in the $d\sigma/dt$
in $pp$ scattering  at large values of $-t$ and hadronic glory
effect would be observed.
The concept of reflective scattering itself is rather general,
 and results from  the unitarity saturation  for $S$-matrix which is
related to  the necessity to reproduce a total cross section growth at $s\to\infty$.

 \begin{theacknowledgments}
  Authors are  grateful to N. Buttimore, L. Jenkovszky, U. Maor, A. Martin,
  V. Petrov and O. Selyugin for
  interesting discussions. One of the authors (S.T.) is also grateful to the Organizers of
  Diffraction 2008, in particular, to A. Bravar, R. Fiore, A. Papa and J. Soffer for
   support and
   warm hospitality at La Londe-les-Maures.
\end{theacknowledgments}

\end{document}